# Novel Reversible 'TSG' Gate and Its Application for Designing Components of Primitive Reversible/Quantum ALU

Himanshu Thapliyal
Center for VLSI and Embedded System Technologies
International Institute of Information Technology
Hyderabad-500019, India
thapliyalhimanshu@yahoo.com

M.B Srinivas
Center for VLSI and Embedded System Technologies
International Institute of Information Technology
Hyderabad-500019, India
srinivas@iiit.net

*Abstract*— In recent years, reversible logic has emerged as a promising computing paradigm having application in low power CMOS, quantum computing, nanotechnology, and optical computing. The classical set of gates such as AND, OR, and EXOR are not reversible. This paper utilizes a new 4 * 4 reversible gate called TSG gate to build the components of a primitive reversible/quantum ALU. The most significant aspect of the TSG gate is that it can work singly as a reversible full adder, that is reversible full adder can now be implemented with a single gate only. A Novel reversible 4:2 compressor is also designed from the TSG gate which is later used to design a novel 8x8 reversible Wallace tree multiplier. It is proved that the adder, 4:2 compressor and multiplier architectures designed using the TSG gate are better than their counterparts available in literature, in terms of number of reversible gates and garbage outputs. This is perhaps, the first attempt to design a reversible 4:2 compressor and a reversible Wallace tree multiplier as far as existing literature and our knowledge is concerned. Thus, this paper provides an initial threshold to build more complex systems which can execute complicated operations using reversible logic.

## I. INTRODUCTION

This section provides an effective background of reversible logic with its definition and the motivation behind it.

### A. Definitions

Researchers like Landauer have shown that for irreversible logic computations, each bit of information lost generates kTln2 joules of heat energy, where k is Boltzmann's constant and T the absolute temperature at which computation is performed [1]. Bennett showed that kTln2 energy dissipation would not occur, if a computation is carried out in a reversible way [2], since the amount of energy dissipated in a system bears a direct relationship to the number of bits erased during computation. Furthermore, voltage-coded logic signals have energy of $E_{sig}$ = ½$CV^2$, and this energy gets dissipated whenever switching occurs in conventional (irreversible) logic implemented in modern CMOS technology. It has been shown that reversible logic helps in saving this energy using charge recovery process [22]. Reversible circuits are those circuits that do not lose information. Reversible computation in a system can be performed only when the system comprises of reversible gates. These circuits can generate unique output vector from each input vector, and vice versa, that is, there is a one-to-one mapping between input and output vectors. Thus, an NXN reversible gate can be represented as

Iv=(I1,I2,I3,I4,……………………IN)

Ov=(O1,O2,O3,…………………ON).

Where Iv and Ov represent the input and output vectors respectively. Classical logic gates are irreversible since input vector states cannot be uniquely reconstructed from the output vector states. There is a number of existing reversible gates such as Fredkin gate[3,4,5], Toffoli Gate (TG) [3, 4] and New Gate (NG) [6].

### B. Motivation behind Reversible Logic

The reversible logic operations do not erase (lose) information and dissipate very less heat. Thus, reversible logic is likely to be in demand in high speed power aware circuits. Reversible circuits are of high interest in low-power CMOS design [10], optical computing [11], quantum computing [12] and nanotechnology [13]. The most prominent application of reversible logic lies in quantum computers. A quantum computer can be viewed as a quantum network (or a family of quantum networks) composed of quantum logic gates; each gate performs an elementary unitary operation on one, two or more two–state quantum systems called qubits. Each qubit represents an elementary unit of information corresponding to the classical bit values 0 and 1. Any unitary operation is reversible, hence quantum networks effecting elementary arithmetic operations such as addition, multiplication and exponentiation cannot be directly deduced from their classical Boolean counterparts (classical logic gates such as AND or OR are clearly irreversible).Thus, Quantum Arithmetic must be built from reversible logic components [14].

### C. Proposed Contribution In Reversible Logic

In this paper, the focus is on the application of a new reversible 4*4 TSG gate [23,24]. The most significant aspect of the proposed gate is that it can work singly as a reversible full adder, that is reversible full adder can now be implemented with a single gate only. A novel reversible 4:2 compressor is also designed from the proposed TSG gate. Furthermore, the optimized adder and 4:2 compressor are used to design the novel 8x8 reversible Wallace tree multiplier. It is proved that the adder, 4:2 compressor and multiplier architecture designed using the proposed TSG gate are better than their counterparts existing in literature, in terms of number of reversible gates and garbage outputs. This is the first attempt to design a reversible 4:2 compressor and reversible Wallace tree multiplier as far as existing literature and our knowledge is concerned. Minimizing the number of garbage outputs and reversible gates is one of the major concerns in reversible logic [7,8,9] and it is observed that TSG gate achieves this leading to high speed and low power reversible circuits. The reversible circuits proposed and designed in this work form the basis for an ALU of a primitive quantum CPU.

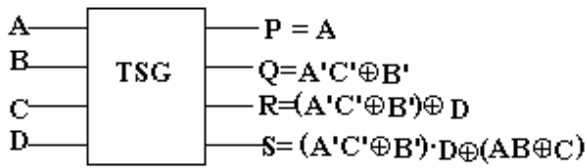

Figure 1. Proposed TSG Gate

## II. PROPOSED 4*4 REVERSIBLE GATE

The authors recently proposed a 4*4 one through reversible gate called TS gate (TSG) [23,24] which is shown in Figure 1. It can be verified that input pattern corresponding to a particular output pattern can be uniquely determined. The proposed TSG gate is capable of implementing all Boolean functions and can also work singly as a reversible Full adder. Figure 2 shows the implementation of the proposed gate as a reversible Full adder.

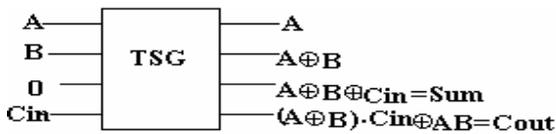

Figure 2. TSG gate Implemented as a Full Adder

A number of reversible full adders were proposed in [6,7,8,9]. The reversible full adder circuit in [6] requires three reversible gates (two 3*3 new gates and one 2*2 Feynman gate) and produces three garbage outputs [garbage output refers to the output that is not used for further computations. In other words, it is not used as a primary output or as an input to other gate]. The reversible full adder circuit in [7,8] requires three reversible gates (one 3*3 new gate, one 3*3 Toffoli gate and one 2*2 Feynman gate) and produces two garbage outputs. The design in [9] requires five reversible Fredkin gates and produces five garbage outputs. The proposed full adder using TSG in Figure 2 requires only one reversible gate (one TSG gate) and produces only two garbage outputs. Hence, the full-adder design in Figure 2 using TSG gate is better than the previous full-adder designs of [6,7,8,9]. A comparison of various full adders is shown in Table I.

TABLE I. A COMPARISON OF VARIOUS REVERSIBLE FULL-ADDER CIRCUITS

| | Number of Gates | Number of Garbage Outputs | Unit Delay |
|---|---|---|---|
| Proposed Circuit | 1 | 2 | 1 |
| Existing Circuit[6] | 3 | 3 | 3 |
| Existing Circuit [7,8] | 3 | 2 | 3 |
| Existing Circuit[9] | 5 | 5 | 5 |

## III. REVERSIBLE 4:2 COMPRESSOR

The 4:2 compressor was introduced by Weinberger in 1981 which he called "4-2 carry save module"[16]. The 4:2 compressor structure actually compresses five partial products bits into three in which four of the inputs are coming from the same bit position of the weight j while one bit is fed from the neighboring position j-1(known as carry-in). The outputs of 4:2 compressor consists of one bit in the position j and two bits in the position j+1. The structure of 4:2 compressor is shown in Figure 3. The efficiency of such a structure is higher since it reduces the number of partial product bits by one half at each stage. The delay of 4:2 compressor is 3 XOR gates in series making it more efficient than using 3:2 counters (full adder) in a regular *"Wallace Tree"* and has important feature that the interconnections between 4:2 cells follow more regular pattern than in case of the *"Wallace Tree"*[17,18,25].

In this paper, a reversible 4:2 compressor is also proposed as shown in Figure 4. The reversible 4:2 compressor is designed from two TSG gates, and it appears from the survey of literature that this is first attempt to design reversible 4:2 compressor. Figure 5 shows the block diagram of 4:2 compressor. Table II shows the comparison results of the 4:2 compressor using TSG with its implementation using existing reversible full adders.

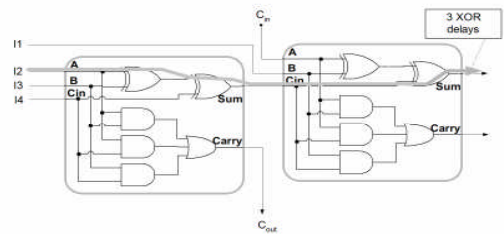

Figure 3. Logic Diagram of 4:2 Compressor [25]

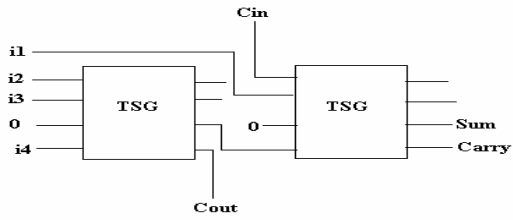

Figure 4.  Proposed Reversible 4:2 Compressor Designed from TSG gate

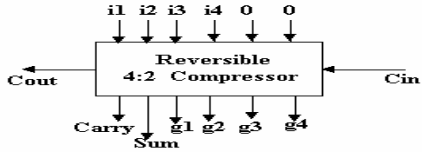

Figure 5.  Block Diagram of Reversible 4:2 Compressor

TABLE II.  A COMPARISON OF 4:2 COMPRESSORS USING VARIOUS FULL ADDER CIRCUITS

|  | Number of Reversible Gates | Number of Garbage Outputs | Unit Clock Cycle |
|---|---|---|---|
| Full adder Using TSG | 2 | 4 | 2 |
| Existing Circuit[6] | 6 | 6 | 6 |
| Existing Circuit [7,8] | 6 | 4 | 4 |
| Existing Circuit[9] | 10 | 10 | 10 |

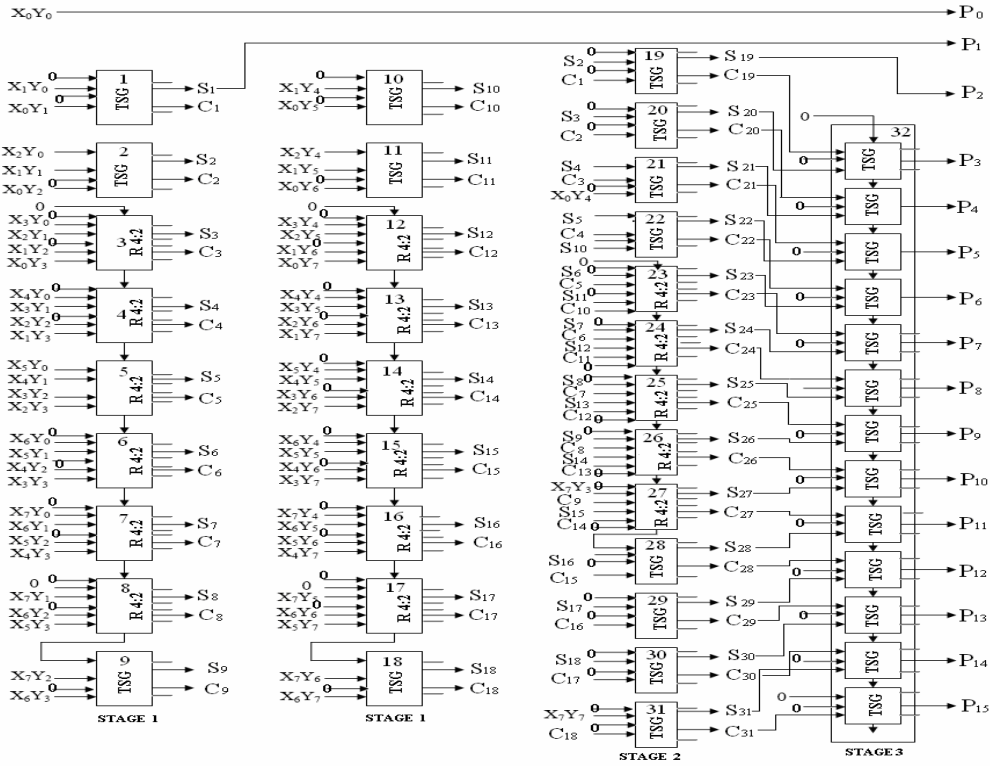

(b)

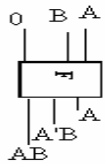

(a)

Figure 6.  (a) Use of Fredkin Gates to Generate the Partial Products In Parallel
(b) Proposed Reversible 8x8 Wallace Tree Multiplier Using Reversible 4:2 Compressor(R 4:2) and TSG gates

## IV. REVERSIBLE WALLACE TREE

A multiply operation consists of three stages: partial products generation stage, partial products addition stage, and the final addition stage [19,20,21]. The second stage is the most important and determines the overall speed of the multiplier. In high-speed designs, the Wallace tree construction method is usually preferred to add the partial products in a tree-like fashion. The Wallace tree is fast since the critical path delay is proportional to the logarithm of the number of bits in the multiplier.

### A. Reversible Wallace Tree Construction

The method used to construct the Wallace tree considers all the bits in each fours rows at a time and compress them in an appropriate manner. The Wallace tree uses 4:2 compressor, full adders and half adders to compress the partial products tree. This paper proposes a reversible 8x8 Wallace tree multiplier using the reversible TSG gate. The proposed architecture can be generalized for NXN bits. The reversible half adder, full adder and 4:2 compressor designed from reversible TSG gate are used as the basic building blocks for the design of reversible 8x8 Wallace tree multiplier. The generation of all partial products of the multiplication can be done in parallel by using Fredkin gates (for ANDing the bits of the multiplier and multiplicand) as shown in Figure 6-a. Then the addition of partial products is performed using reversible Wallace tree. The Wallace tree uses reversible 4:2 compressors, full adders and half adders designed from TSG gate. Figure 6-b shows the proposed reversible 8x8 Wallace tree multiplier in which (R 4:2) represents reversible 4:2 compressor. In Stage 1, the partial products are added using 4:2 compressors, full adders and half adders, the S and C generated of all the blocks are arranged according to their weights. In the Stage 2, the reduced partial products are again added using 4:2 compressors, full adders and half adders. Finally, the result obtained in Stage 2 is finally added using a parallel adder designed from TSG gates to generate the product bits P0…P14, P15.

### B. Evaluation of the Proposed Wallace Tree Multiplier

The proposed reversible 8x8 Wallace Tree is designed from reversible TSG gate. This is the first attempt in literature to design the reversible 8x8 Wallace Tree multiplier. The novelty lies in the introduction of reversible TSG gate and its implementation for designing the adder and compressor, which are later used to design the reversible Wallace tree multiplier. The proposed reversible architecture of the multiplier can be improved further by implementing block 32 in the architecture (Figure 6-b) with the efficient parallel adder architectures proposed in [23-24]. There seems to be no existing counterpart in literature and hence no comparative study is done. It has already been proved in the earlier sections that the adder and compressor designed from the TSG gate are the most optimized one, thus making the overall architecture of the Wallace tree multiplier as the most optimal one.

## V. CONCLUSIONS

The focus of this paper is the application of a new reversible 4*4 TSG gate to design the components of a primitive quantum/reversible ALU. The novel optimized adder and 4:2 compressor are designed from TSG gate, which are later used to design the novel 8x8 reversible Wallace tree multiplier. It is proved that the adder, 4:2 compressor and multiplier architectures designed with the proposed TSG gate are better than their existing counterparts, in terms of number of reversible gates and garbage outputs, leading to a high speed and low power reversible circuits. All the proposed architectures are analyzed in terms of technology independent implementations. The proposed circuit can be used for designing large reversible systems. Thus, this paper provides the initial threshold to build more complex reversible systems.